\documentclass[journal]{IEEEtran}

\usepackage{amsfonts}
\usepackage{amsmath}
\usepackage{amssymb}
\usepackage{graphicx}
\usepackage{mathtools}
\usepackage{adjustbox}
\usepackage{hyperref}
\hypersetup{colorlinks,allcolors=black}
\usepackage{comment}
\usepackage{cite} 

\usepackage[caption=false, font=footnotesize]{subfig} 

\usepackage{amsthm} 

\usepackage{soul}  

\usepackage[normalem]{ulem}
\newtheorem{theorem}{Theorem}

\usepackage{hyperref}

\usepackage[makeroom]{cancel} 

\usepackage{xcolor} 

\usepackage{algorithm}
\usepackage{algorithmicx}
\usepackage{algpseudocode}

\usepackage{tikz}
\usetikzlibrary{shapes.multipart}

\usetikzlibrary{circuits.logic.US, positioning}
\usetikzlibrary{shapes.geometric, arrows}

\begin{document}
\title{Efficient Low-Memory Fast Stack Decoding with Variance Polarization for PAC Codes}
\author{Mohsen~Moradi\textsuperscript{\href{https://orcid.org/0000-0001-7026-0682}{\includegraphics[scale=0.06]{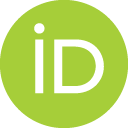}}},
Hessam~Mahdavifar\textsuperscript{\href{https://orcid.org/0000-0001-9021-1992}{ \includegraphics[scale=0.06]{figs/ORCID}}}
\thanks{Mohsen Moradi is with the School of Electrical, Computer and Energy Engineering, Arizona State University, Tempe, AZ 85287, USA. (e-mail: mmorad11@asu.edu).}
\thanks{Hessam Mahdavifar is with the Department of Electrical \& Computer Engineering, Northeastern University, Boston MA-02115, USA (e-mail: h.mahdavifar@northeastern.edu).}
\thanks{This work was supported by NSF under Grant CCF-2415440 and the Center for Ubiquitous Connectivity (CUbiC) under the JUMP 2.0 program.}%
}

\maketitle

\begin{abstract}
Polarization-adjusted convolutional (PAC) codes have recently emerged as a promising class of error-correcting codes, achieving near-capacity performance particularly in the short block-length regime. In this paper, we propose an enhanced stack decoding algorithm for PAC codes that significantly improves parallelization by exploiting specialized bit nodes, such as rate-0 and rate-1 nodes. For a rate-1 node with $N_0$ leaf nodes in its corresponding subtree, conventional stack decoding must either explore all $2^{N_0}$ paths, or, same as in fast list decoding, restrict attention to a constant number of candidate paths. 
In contrast, our approach introduces a pruning technique that removes candidate paths with small path metrics while ensuring that the probability of pruning the correct path decays exponentially with the threshold.
Furthermore, we propose a novel approximation method for estimating variance polarization under the binary-input additive white Gaussian noise (BI-AWGN) channel. Leveraging these approximations, we develop an efficient stack-pruning strategy that selectively preserves decoding paths whose bit-metric values align with their expected means. This targeted pruning substantially reduces the number of active paths in the stack, thereby decreasing both decoding latency and computational complexity. Numerical results demonstrate that for a PAC$(128,64)$ code, our method achieves up to a 70\% reduction in the average number of paths without degrading error-correction performance.
\end{abstract}


\maketitle

\section{Introduction}
\IEEEPARstart{P}{olar} codes are the first class of error-correcting codes that theoretically achieve the capacity of memoryless symmetric channels \cite{arikan2009channel}. 
However, at short block lengths, their error-correction performance (ECP) falls short of the theoretical limits. 
The recently introduced polarization-adjusted convolutional (PAC) codes \cite{arikan2019sequential}, which enhance polar codes via pre-coding with convolutional codes, exhibit an ECP that closely approaches the theoretical bounds at short block length regimes when decoded using sequential decoding algorithms \cite{arikan2019sequential, moradi2020performance}.
PAC codes can also be viewed as a combination of polar codes and cyclic codes \cite{moradi2024polarization}.
PAC codes can be decoded using various tree-search algorithms, such as sequential decoding.

As a sequential decoding algorithm, the stack algorithm was first introduced for convolutional codes \cite{zigangirov1966some,jelinek1969fast}. At each step, it selects the most likely path in the stack, extends it along both branches, computes the corresponding path metrics, and inserts the resulting paths into the stack after removing the original path.
Stack decoding and related sequential tree-search methods have also been studied for both polar and PAC codes \cite{Niu2012StackPolar,Song2016SegmentedSCS,Xiang2019LogSCS,Minja2021BidirectionalStack,Timokhin2023SCCreeper,Moradi2022BitFlippingStackPAC,Rowshan2021PACSequentialVsList}.
Because the stack algorithm may revisit the same decoding levels multiple times, extending both branches at each step can significantly increase the number of candidate paths in the stack, which complicates the search for the most likely path and increases decoding delay.
In this paper, we propose a stack-pruning method that identifies branches unlikely to lie on the correct path and prevents them from being inserted into the stack.
Our results show that the proposed pruning approach significantly reduces the average stack size, thereby reducing both decoding complexity and delay.

Our stack-pruning technique benefits from the concept of metric function polarization introduced in \cite{moradi2024fast}, where it is shown that at each step of the polarization process, the mean and variance of the metric function random variable correspond to the polarized mutual information and varentropy (variance) of the synthesized binary input discrete memoryless channel (BI-DMC), respectively. 
For a binary input additive white Gaussian noise (BI-AWGN) channel, \cite{tal2013construct} proposes an algorithm to calculate the polarized mutual information, which can also be used to compute the polarized channel varentropy \cite{moradi2024fast}.
In \cite{moradi2024fast}, also an approximation method for calculating the varentropy of the BI-AWGN channel is introduced.
Following the approach in \cite{li2013practical}, in this paper, we also assume that the channel remains a BI-AWGN channel after each polarization step. Using the method outlined in \cite{moradi2024fast}, we propose a technique to compute the polarized varentropies of the synthesized BI-AWGN channels at each step of the polarization.
Numerical results show that the polarized varentropy values obtained from our approximation method closely match those derived using the method from \cite{tal2013construct}.

Using our proposed approximation for the polarized mutual information and polarized varentropy of the BI-AWGN channel, we introduce a stack decoding method that prunes branches falling below a certain threshold. 
Our stack-pruning method integrates the approaches proposed in \cite{moradi2023tree} and \cite{moradi2024fast}.
In \cite{moradi2023tree}, it was shown using the Chernoff bound that the probability of pruning the correct path, i.e., the probability that the correct path's metric falls below the threshold, decreases exponentially with the threshold in the exponent. Similarly, in \cite{moradi2024fast}, using the Chebyshev bound, it was demonstrated that the probability of the correct path’s metric deviating from the polarized mutual information by more than the threshold is upper bounded by a ratio, with the polarized channel varentropy in the numerator and the square of the threshold in the denominator.

After polarization, we observe that for certain polarized channels, the Chebyshev bound provides a tighter threshold, while for others, the Chernoff bound yields a better result. Leveraging both bounds, along with our proposed polarized varentropy approximation, we introduce a pruning method that excludes paths with metric values below the underlying threshold, which is determined based on a suitable method out of the two methods mentioned above.
Numerical results demonstrate that for a PAC$(128,64)$ code, our proposed pruning method achieves a 70\% reduction in the average number of paths in the stack compared to the conventional stack algorithm, and an 18\% improvement compared to the method in \cite{moradi2023tree} at $1~$dB $E_b/N_0$.

To enhance the parallelism of the proposed fast stack decoder, we consider four types of special nodes: rate-0, rate-1, type-IV, and REP nodes. Other special nodes, such as single-parity-check (SPC) \cite{zhu2023fast, yin2024software, ji2023low} and Type-T nodes \cite{dai2024fast}, are also of interest, and extending the proposed framework to incorporate them is left for future work.
A rate-1 node with \( N_0 \) leaf nodes can generate \( 2^{N_0} \) candidate paths in the conventional stack decoding algorithm. In \cite{trofimiuk2020fast}, a fast block stack decoding method for polar codes is introduced, which recommends retaining only 4 out of the \( 2^{N_0} \) possible paths to avoid significant ECP loss.
In polar codes, the rate profile is typically constructed based on the reliability of the polarized bit-channels, and retaining 4 paths is generally sufficient to preserve ECP.
However, for short block-length PAC codes, a rate profile based on alternative criteria—such as the weight distribution of the code—is often more favorable \cite{moradi2020performance}.
In our approach, for rate-1 nodes, we apply a stack-pruning strategy that permits bifurcation only when the corresponding child node has a bit metric exceeding our proposed threshold. 
For REP nodes, which yield two candidate paths, we retain a path only if the bit metric of its final leaf node exceeds its associated threshold.
Our proposed algorithm for handling rate-IV and REP special nodes can also be seamlessly applied to any other rate profile regardless of the particular strategy to select information-bit positions.
Such nodes are investigated in the context of successive-cancellation list (SCL) decoding in \cite{yao2025low}.
In particular, since the pruning and metric-update rules we introduce do not depend on a particular rate profile structure, the method remains applicable even when the bit-channel ordering or information set is tailored to alternative design criteria.
Our contributions in this paper can be summarized as follows:

\begin{itemize}
    \item \textbf{Approximation of Polarized Varentropy:} We develop an efficient approximation for estimating the polarized channel varentropy of a BI-AWGN channel in $O(\log_2(N))$ time. This approximation enables fast and accurate computation of adaptive pruning thresholds without compromising decoding performance.
    
    \item \textbf{Hybrid Path-Pruning Strategy:} We propose a novel pruning method that exploits the polarization properties of the bit-metric function to discard low-probability decoding paths while guaranteeing that the probability of discarding the correct path decreases exponentially. By combining insights from both Chebyshev and Chernoff bounds, our method dynamically selects the tighter threshold for each bit-channel, resulting in a more effective and adaptive pruning method.
    
    \item \textbf{Fast Stack Decoding Algorithm for PAC Codes:} We design an efficient fast stack decoding algorithm for PAC codes that seamlessly integrates the proposed pruning strategy and supports the handling of rate-0, REP, type-IV, and rate-1 special nodes. For rate-1 nodes, where conventional stack decoding may require exploring up to $2^{N_0}$ candidate paths—which can be prohibitively large for high-rate PAC codes—our pruning-based approach significantly reduces computational complexity while maintaining near-optimal error-correction performance.
    
\end{itemize}

The remainder of this paper is organized as follows. 
Section~\ref{Sec: Preliminaries} reviews the necessary background and preliminaries to make the paper self-contained. 
In Section~\ref{sec: MetricPolarization}, we discuss the polarization of channel parameters and introduce an approximation method for estimating the polarized varentropy. 
Section~\ref{sec:pruning} presents our proposed pruning strategy, which selectively removes unlikely decoding paths from the stack based on adaptive thresholds. 
In Section~\ref{sec:FastStack}, we introduce the fast stack decoding algorithm, which efficiently handles different types of special nodes and integrates the proposed pruning method to reduce decoding complexity. 
Section~\ref{sec:Numerical} provides numerical results demonstrating the performance gains and computational benefits of the proposed techniques.
Finally, Section~\ref{sec: Conclusion} concludes the paper.


\section{Preliminaries} \label{Sec: Preliminaries}

\subsection{Notation Convention}

Vectors and matrices are represented in this paper by boldface letters. 
All operations are performed in the binary field $\mathbb{F}_2$. 
For a vector $\mathbf{u} = (u_1, u_2,\ldots,u_N)$,
subvectors $(u_1, \ldots, u_i)$ and $(u_i, \ldots, u_j)$ are denoted by $\mathbf{u}^i$ and $\mathbf{u}_i^j$, respectively.
For a given set $\mathcal{A}$, the subvector $\mathbf{u}_{\mathcal{A}}$ includes all elements $u_i$ where $i$ belongs to the set $\mathcal{A}.$


\subsection{PAC Coding Scheme}  \label{sec: PAC_BG}

A PAC code can be characterized by the parameters $(N, K, \mathcal{A}, \mathbf{c}(t))$, where $N = 2^n$ is the code length, $K$ is the data length, $\mathcal{A}$ defines the location of the data bits (known as the rate profile), and $\mathbf{c}(t)$ is the connection polynomial that specifies the convolutional code within the PAC code. 
During encoding, the data vector $\mathbf{d}$ is first embedded into the data carrier vector $\mathbf{v}$ such that $\mathbf{v}_{\mathcal{A}} = \mathbf{d}$ and $\mathbf{v}_{\mathcal{A}^c} = \mathbf{0}$. A one-to-one convolutional encoder, derived from the connection polynomial, then maps the data carrier vector $\mathbf{v}$ to the vector $\mathbf{u}$. 
Subsequently, the polar transform maps $\mathbf{u}$ to the encoder output $\mathbf{x}$ as $\mathbf{x} = \mathbf{u}\mathbf{F}^{\otimes n}$, where $\mathbf{F}=\bigl[\begin{smallmatrix}1 & 0 \\ 1 & 1\end{smallmatrix}\bigr]$ and $\otimes$ denotes the Kronecker power.
With binary phase shift keying (BPSK) modulation over a BI-AWGN channel, the channel output is represented by the vector $\mathbf{y}$. A stack decoder is employed to estimate the vector $\hat{\mathbf{v}}$, from which the estimate of the data vector is retrieved using the rate profile.
The frame error rate (FER), defined as \( P(\hat{\mathbf{d}} \ne \mathbf{d}) \), serves as the key metric to evaluate the system performance.

\subsection{Rate Profile}

For short-blocklength PAC codes, the choice of the rate profile $\mathcal{A}$ has a significant impact on both the ECP and the decoding complexity.
As shown in \cite{arikan2019sequential, moradi2020performance}, PAC codes with a Reed–Muller (RM)-based rate profile can achieve performance close to theoretical bounds for certain code lengths and code rates.
In an RM-based construction, the information set $\mathcal{A}$ is selected according to the rows of $\mathbf{F}^{\otimes n}$ with the largest Hamming weights.
A limitation of a pure RM-based construction is that it is not possible for designing PAC codes at arbitrary rates.
To obtain rate profiles for a broader range of rates, alternative construction methods can be employed.
One approach is to use a genetic-algorithm-based optimization initialized from the RM profile, which enables the design of rate profiles for additional rates \cite{moradi2024pac_Genetic}.
Another flexible approach is Monte-Carlo (MC)-based construction, which can be used to design rate profiles for essentially arbitrary rates \cite{moradi2025monte}.
Reinforcement-learning-based methods (Q-learning) have also been proposed for rate-profile optimization \cite{mishra2022modified}.
In this work, the proposed fast stack decoding and pruning framework is not restricted to a specific rate-profile construction.
Once the information set $\mathcal{A}$ is specified, the same decoding algorithm can be applied to RM-based, Q-learning, genetic-algorithm-based, and MC-based rate profiles.


\subsection{Channel Parameters}
For a BI-DMC $W: \mathcal{X} \longrightarrow \mathcal{Y}$, where $\mathcal{Y}$ represents the output alphabet, the channel transition probability is denoted by $W(y|x)$, with $y \in \mathcal{Y}$ and $x \in \mathcal{X} = \{0,1\}$. The channel cutoff rate for a uniform input distribution is defined as
\begin{equation}\label{eq:R0vZW}
   R_0(W) \triangleq 1 - \log_2 \left( 1+ \sum_{y \in \mathcal{Y}} \sqrt{W(y|0)W(y|1)}\right). 
\end{equation}
Another important parameter of the channel $W$ is the symmetric channel capacity, defined as
\begin{equation}
    I(W)  \triangleq \sum_{x \in \mathcal{X}} \sum_{y \in \mathcal{Y}} p(x,y) \log_2 \left(\frac{p(y|x)}{p(y)} \right).
\end{equation}
The varentropy of the channel $W$ is also defined as
\begin{equation}
    \text{Var}(W) \triangleq \sum_{x \in \mathcal{X}} \sum_{y \in \mathcal{Y}} p(x,y) \log_2^2 \left(\frac{p(y|x)}{p(y)} \right) - I(W)^2.
\end{equation}

\subsection{Stack Decoding}
In a sequential decoding algorithm, the decoder bifurcates only the best path, i.e., the path with the best path metric value. As a sequential decoding algorithm, the stack algorithm maintains all previously evaluated paths in a stack. For PAC codes, the stack decoding procedure is described in Algorithm \ref{alg:stack}.

\begin{algorithm}[t!]
\footnotesize
\caption{Stack Algorithm for PAC Code}\label{alg:stack}
\begin{algorithmic}[1]
\State \textbf{Initialize} the stack with the single-node path starting from the origin with path metric of $0$.

\While{the top path has not reached a leaf node} \label{loop}
    \State \textbf{Pop} the top path from the stack.
    \If{the current node corresponds to an information bit}
        \State \textbf{Compute} the path metric values for both immediate successors.
    \Else
        \State \textbf{Compute} the path metric value for the single immediate successor.
    \EndIf
    \State \textbf{Push} the successor(s) to the stack with their associated metric values.
    \State \textbf{Reorder} the stack in descending order of the path metric values.
\EndWhile

\State \textbf{Output} the top path as the decoded path.

\end{algorithmic}
\end{algorithm}

\subsection{Path Metric}
For a code of length $N$, after polarization, one can obtain $N$ synthesized channels $W_N^{(i)}: \mathcal{X} \longrightarrow \mathcal{Y}^{N} \times \mathcal{X}^{i-1}$ from $N$ copies of the channel $W$ \cite{arikan2009channel}. 
The path metric for the stack decoding at level $j$ is defined in \cite{moradi2021sequential} as
\begin{equation}
    \Gamma(\mathbf{u}^j; \mathbf{y}) = \sum_{i=1}^{j} \gamma(u_i; \mathbf{y}, \mathbf{u}^{i-1}),
\end{equation}
where
\begin{equation}
\begin{split}
     \gamma(u_i; \mathbf{y}, \mathbf{u}^{i-1}) 
     &= \log_2 \frac{P(\mathbf{y}, \mathbf{u}^{i-1} | u_i)}{P(\mathbf{y}, \mathbf{u}^{i-1})} - R_0(W_N^{(i)}) \\
     &= 1 - \log_2 \left(1 + 2^{-L_i \cdot (-1)^{u_i}  }\right) - R_0(W_N^{(i)}),
\end{split}
\end{equation}
and
\begin{equation}
    L_i = \log_2 \left( \frac{P(\mathbf{y}, \mathbf{u}^{i-1} | u_i = 0)}{P(\mathbf{y}, \mathbf{u}^{i-1} | u_i = 1)} \right).
\end{equation}
The bit-channel cutoff rate $R_0(W_N^{(i)})$ is employed as the bias in the stack decoding algorithm.

\subsection{BI-AWGN}
For a BI-AWGN channel, the channel cutoff rate is given by \cite[p. 142]{viterbi2009principles}
\begin{equation}
    R_0(W) = 1 - \log_2 \left( 1+ e^{\frac{-1}{2\sigma^2}} \right),
\end{equation}
where $\sigma^2$ is the noise variance.

The symmetric capacity $I(W)$ can be approximated as
\begin{equation}\label{eq:J_I}
    J_{\text{approx}}(t) = \left[1 - 2^{-0.3073 \, t^{2 \times 0.8935}} \right]^{1.1064},
\end{equation}
where $t = \frac{2}{\sigma}$ \cite{brannstrom2004convergence}.

Furthermore, the varentropy $V(W)$ can be approximated as \cite{moradi2024fast}
\begin{equation}\label{eq:V-approx}
    V_{\text{approx}}(t) = -(J_{\text{approx}}(t) - 1)^2 - K_{\text{approx}}(t) + 1,
\end{equation}
where 
\begin{equation}\label{eq:Kt_approx}
    K_{\text{approx}}(t) = \left[1 - 2^{-0.96483 \, t^{2 \times 0.61746}} \right]^{10.232}.
\end{equation}

\subsection{Sequential Decoders for PAC Codes}

Stack decoding and Fano decoding are both sequential decoders, but they offer different and practically important tradeoffs. In stack decoding, the decoder always expands the currently most promising partial path (highest
metric) and keeps competing candidates available for later refinement.
This best-first behavior avoids long excursions on weak paths and reduces the sensitivity to step-size/threshold-tuning choices that are inherent in Fano decoding search. In particular, Fano decoding relies on a
threshold rule: when forward progress along the current path fails to meet the threshold, the decoder decreases the threshold and/or backtracks to earlier depths to explore alternative branches. Such backtracking can be
frequent in low-SNR regimes or for unfavorable local metric fluctuations, creating highly variable decoding time and additional latency due to repeated revisits of previously explored tree levels. Moreover, stack decoding provides an explicit
\emph{complexity/latency knob}: by limiting the stack size and the number of cycles, one can enforce a predictable worst-case
budget while retaining near-optimal performance. In our experiments, a bounded stack decoder achieves FER comparable to
Fano decoding even when Fano is allowed to explore an unbounded number of paths (see Fig.~2 and Fig.~3).

Compared to list decoding (e.g., SCL), stack decoding can be viewed as an \emph{adaptive list} method: instead of maintaining
a fixed list of size $L$ at every information bit, the stack keeps only those partial paths that remain competitive under the
path-metric ordering. This is particularly attractive for short-blocklength PAC codes, where high-rate segments (e.g., rate-1
subtrees) can otherwise trigger a large expansion; classical fast-list approaches must restrict the candidate set
to a small constant to control complexity, which may not be robust across PAC rate profiles.
In contrast, our proposed pruning strategy leverages metric polarization/varentropy to admit bifurcations only when the
corresponding bit-metrics are consistent with their expected behavior, reducing the \emph{average} number of active paths
substantially (up to $\approx 70\%$ in our results) without degrading error-correction performance.
Finally, stack decoding admits an efficient implementation via a heap-based data structure with logarithmic-time stack
operations, enabling a low-memory decoder that naturally carries the per-path convolutional state needed for PAC decoding.

\section{Polarization of Channel Parameters}\label{sec: MetricPolarization}
Following the approach in \cite{li2013practical}, we assume that the log-likelihood ratio (LLR) of the channel output $W$ follows a Gaussian distribution with mean $2/\sigma^2$ and variance $4/\sigma^2$, and that after each polarization step, the resulting channels continue to behave as BI-AWGN channels. This allows the computation of $\sigma_N^{(i)}$ after $\log_2(N)$ polarization steps for the $i$th polarized channel. 
Building on this model, in this paper we propose using the cutoff rate of the $i$th bit-channel as
\begin{equation}
    R_0(W_N^{(i)}) = 1 - \log_2 \left( 1 + e^{\tfrac{-1}{2(\sigma_N^{(i)})^2}} \right).
\end{equation}
Similarly, we propose that the mutual information of the polarized $i$th bit-channel, $I(W_N^{(i)})$, be approximated by $J_{\text{approx}}(t)$, and that its varentropy be approximated by $V_{\text{approx}}(t)$, where $t = \frac{2}{\sigma_N^{(i)}}$.

The mean of the $i$th bit-metric function random variable is \cite{moradi2023tree}
\begin{equation}
    \mathbb{E}\left[ \gamma(U_i; \mathbf{Y}, \mathbf{U}^{i-1}) \right] = I(W_N^{(i)}) - R_0(W_N^{(i)}),
\end{equation}
and its variance is \cite{moradi2024fast}
\begin{equation}
    \text{Var}\left( \gamma(U_i; \mathbf{Y}, \mathbf{U}^{i-1}) \right) = V(W_N^{(i)}).
\end{equation}

\section{Stack pruning method}\label{sec:pruning}

The defining idea of sequential decoding is that the decoder focuses only on the most promising candidate paths.
If a path to a given node appears weak in terms of its metric, we can discard all subsequent paths from that node without causing any meaningful degradation in performance compared to maximum likelihood decoding.
The path metric function directs the decoder toward exploring the most promising paths, which naturally brings us to the stack algorithm.

In a PAC code, the convolutional code sees a polarized channel, and most of the synthesized bit channels have very small varentropy, resulting in bit-metric values very close to the polarized channel capacity. 
In addition to using the path metric function values to order the paths in the stack, we leverage the varentropy polarization by discarding branches corresponding to an information bit, along with all their descendant paths, if their bit-metric values deviate significantly from their mean.

Using Chebyshev's inequality, the following theorem from \cite{moradi2024fast} demonstrates that the bit-metric random variable of a correct path tends to concentrate around the bit-channel capacity in both nearly noiseless and highly noisy channels.

\begin{theorem}\cite{moradi2024fast}.\label{thm:UpperBound}
In a BI-DMC, the bit-metric random variable $\gamma(U_i; \mathbf{Y}, \mathbf{U}^{i-1})$ converges to its expected value in probability. More precisely, it holds that
\begin{equation}\label{eq:Chebyshev}
P\left( \left| \gamma(U_i; \mathbf{Y}, \mathbf{U}^{i-1}) - I(W_N^{(i)}) \right| \geq m_i \right)
 \leq \frac{\text{Var}(W_N^{(i)})}{m_i^2},
\end{equation}
where $m_i$ is a positive real number.
\end{theorem}

The following theorem from \cite{moradi2024fast} also proves that the upper bound of \eqref{eq:Chebyshev} converges to zero.

\begin{theorem}\cite{moradi2024fast}.\label{thm:Var}
In a BI-DMC, the variances of the bit metric random variables approach zero a.e.
\end{theorem}

Note that, in proving Theorem~\ref{thm:Var}, the blocklength must tend to infinity.
In that asymptotic regime, the upper bound in \eqref{eq:Chebyshev} converges to zero a.e.
However, PAC codes are mainly studied in the short-blocklength regime, where the variances of the polarized bit-channels are not yet fully polarized.
In contrast, the bound in Theorem~\ref{thm:UpperBound} is non-asymptotic and is therefore useful for short blocklengths, providing more insight into the finite-length behavior than Theorem~\ref{thm:Var}.

Moreover, using Chernoff's bound, the following theorem from \cite{moradi2023tree} shows that the probability of the bit-metric random variable deviating significantly from its expected value decays exponentially to zero.
Similarly, this theorem is also useful for short-blocklength PAC codes.

\begin{theorem}\cite{moradi2023tree}.\label{thm:Chernof}
Let $\gamma(U_i;\textbf{Y}, \mathbf{U}^{i-1})$ represent the bit-metric random variable. For any constant threshold $m_T$ where $m_T < I(W_N^{(i)}) - R_0(W_N^{(i)})$, the probability that $\gamma(U_i; \mathbf{Y}, \mathbf{U}^{i-1})$ falls below $m_T$ is bounded by
\begin{equation}\label{eq:Chernof}
    P\left\{\gamma(U_i;\mathbf{Y}, \mathbf{U}^{i-1}) \leq m_T \right\} \leq 2^{\frac{m_T}{2}}.
\end{equation}
\end{theorem}

Utilizing the upper bounds from \eqref{eq:Chebyshev} and \eqref{eq:Chernof}, we propose a path-pruning strategy for the stack decoding algorithm as follows. Let $P_{\mathrm{th}}$ denote a design parameter specifying the target upper bound on the probability of pruning the correct branch. By inverting the Chebyshev and Chernoff bounds, we obtain two candidate thresholds for the $i$th bit-channel:
\[
\gamma_{T,\mathrm{Cheb}}^{(i)} = \left\lfloor -\sqrt{\frac{\mathrm{Var}(W_N^{(i)})}{P_{\mathrm{th}}}} \right\rfloor - 10,
~~
\gamma_{T,\mathrm{Cher}}^{(i)} = \left\lfloor \log_2(P_{\mathrm{th}}) \right\rfloor.
\]
We then define a pruning threshold vector $\boldsymbol{\gamma}_T$, whose $i$th element is
\[
\boldsymbol{\gamma}_T^{(i)} =
\min\left\{
\gamma_{T,\mathrm{Cheb}}^{(i)},
\gamma_{T,\mathrm{Cher}}^{(i)}
\right\}.
\]
Accordingly, if the bit-metric value $\gamma(u_i;\mathbf{y},\mathbf{u}^{i-1})$ corresponding to a data bit is less than $\boldsymbol{\gamma}_T^{(i)}$, then the corresponding branch is not explored and is not inserted into the stack.
We use an additional constant offset in the Chebyshev-based threshold as an empirical safety margin.

\section{Fast stack algorithm}\label{sec:FastStack}

We employ a heap data structure \cite{cormen2009introduction,shalvi2011signal}, which enables logarithmic-time operations, to implement a memory-efficient stack decoding algorithm for PAC codes. In this structure, partial path metrics serve as the keys, while the satellite data consists of four components: the convolutional code’s input vector, output vector, and state; and the polar demapper’s intermediate LLRs. The intermediate LLRs and the convolutional code’s input/output vectors are stored in arrays of length $N-1$ and $N$, respectively. For an alternative implementation that stores checksum values (instead of the convolutional code’s outputs) in a vector of length $N/2$, see \cite{moradi2022performance}.

\algnewcommand\Input{\Statex \textbf{Input:} }
\algnewcommand\Output{\Statex \textbf{Output:} }


\begin{algorithm}[t!]
\small
\caption{$\textsc{PAC-Fast-Stack-Decoder}$}
\label{alg: pac_main_decoding}
\begin{algorithmic}[1]

\Input $\mathbf{L}_{ch},\mathcal{A},\mathbf{c},\mathbf{E}_0,\textit{max-cycles},\textit{S-size},\boldsymbol{\gamma}_T$
\Output $\hat{\mathbf{v}}$, $\textit{cycles}$, $S\textit{-used}$

\State $S\_values \gets \textsc{Calculate-S-Values}(\mathbf{RP})$
\State $N \gets \text{length}(\mathbf{L}_{ch})$ \Comment{depth of the tree}

\State $S[1].m \gets 0$
\State $S[1].\mathbf{v} \gets \text{NIL}$
\State $S[1].\mathbf{st}[1,\cdots,|\mathbf{c}|-1] \gets [0,\cdots,0]$
\State $S[1].\mathbf{L}[1, \cdots, N-1] \gets [0, \cdots, 0]$
\State $S[1].\mathbf{u}[1, \cdots, N] \gets [0, \cdots, 0]$

\State $[S_N, \textit{cycles}, n] \gets [1, 0,\log_2(N)]$
\While{$\textit{True}$}
    \If{$\textit{cycles} = \textit{max-cycles}$}
        \State \textbf{break}
    \Else
        \State $\textit{cycles} \gets \textit{cycles} + 1$
    \EndIf
    \State $[\textit{topS}, S, S_N] \gets \textsc{Extract-Max}(S, S_N)$
    \State $\mathbf{st} \gets \textit{topS}.\mathbf{st}$
    \State $i \gets \text{length}(\textit{topS}.\mathbf{v}) + 1$
    \State $[\mathbf{L}, \mathbf{idx}_{\text{LLR}}, \textit{Type}, \mathbf{idx}_{\text{chunk}}, \mathbf{u}] \gets$
    \Statex \hspace{1cm}$\textsc{UpdateLLR}(\mathbf{L}_{ch}, \textit{topS}.\mathbf{L}, i-1, \textit{topS}.\mathbf{u}, n, S\_values)$

    \State $\mathbf{r}_i \gets \mathbf{L}[\mathbf{idx}_{\text{LLR}}]$
    \State $N_{\text{chunk}} \gets \text{length}(\mathbf{idx}_{\text{chunk}})$

    \If{$\textit{Type} = 0$}
        \State $[S, S_N] \gets 
    \textsc{f-Type-0}(\textit{topS}, S, S_N, \mathbf{st}, \mathbf{L}, \mathbf{idx}_{\text{LLR}},$
        \Statex \hspace{3.6cm}$\mathbf{idx}_{\text{chunk}}, \mathbf{u}, \mathbf{c}, \mathbf{E}_0, \text{S-Size}, \mathbf{r}_i, N_{\text{chunk}})$

    \ElsIf{$\textit{Type} = \text{REP}$}
    \State $[S, S_N] \gets \textsc{f-Type-REP}(\textit{topS}, S, S_N, \mathbf{st}, \mathbf{L}, \mathbf{idx}_{\text{LLR}},  $
    \Statex \hspace{2.4cm}$\mathbf{idx}_{\text{chunk}},\mathbf{u},\mathbf{c}, \mathbf{E}_0, \mathbf{RP}, \boldsymbol{\gamma}_T, \text{S-Size}, \mathbf{r}_i, N_{\text{chunk}})$
        
            
    \ElsIf{$\textit{Type} = \text{IV}$}
        \State $[S, S_N] \gets \textsc{f-Type-IV}(\textit{topS}, S, S_N, \mathbf{st}, \mathbf{L}, \mathbf{idx}_{\text{LLR}},  $
        \Statex \hspace{2.4cm}$\mathbf{idx}_{\text{chunk}},\mathbf{u}, \mathbf{c}, \mathbf{E}_0, \mathbf{RP}, \boldsymbol{\gamma}_T, \text{S-Size}, \mathbf{r}_i, N_{\text{chunk}})$
            
    \ElsIf{$\textit{Type} = \text{ALL}$}
        \State $[S, S_N] \gets \textsc{f-Type-All}(\textit{topS}, S, S_N, \mathbf{st}, \mathbf{L}, \mathbf{idx}_{\text{LLR}}, $
        \Statex \hspace{2.4cm}$\mathbf{idx}_{\text{chunk}}, \mathbf{u}, \mathbf{c}, \mathbf{E}_0, \boldsymbol{\gamma}_T, \text{S-Size}, \mathbf{r}_i, N_{\text{chunk}})$

    \EndIf

    \If{$\text{length}(S[1].\mathbf{v}) = N$}
        \State \textbf{break}
    \EndIf
\EndWhile
\State $\hat{\mathbf{v}} \gets S[1].\mathbf{v}$
\State $S\textit{-used} \gets S_N$
\State \Return $[\hat{\mathbf{v}}, \textit{cycles}, S\textit{-used}]$
\end{algorithmic}
\end{algorithm}

Algorithm~\ref{alg: pac_main_decoding} presents the main function of our proposed fast stack decoding algorithm for PAC codes. The algorithm operates on the channel output LLRs, denoted by $\mathbf{L}_{ch}$, the data index set $\mathcal{A}$, the binary coefficient vector $\mathbf{c}$ corresponding to the connection polynomial $c(t)$, where the convolutional memory depth is $|\mathbf{c}|-1$, the vector of bit-channel bias values (bit-channel cutoff rates) $\mathbf{E}_0$, the maximum number of decoding cycles, the stack size, and the threshold vector $\boldsymbol{\gamma}_T$, which determines whether a path should be pruned.
The outputs of the algorithm are the data carrier vector $\hat{\mathbf{v}}$, the number of cycles, and the stack size used (i.e., the number of paths stored in the stack).

Based on the rate profile, $S\_values$ is defined as a cell array containing $\log_2(N)$ vectors. 
The first entry consists of a single value representing the total number of information bits $K$. 
The second entry is a vector of length two, where the first element denotes the number of information bits $K^-$ in the first half of the codeword and the second element denotes the number of information bits $K^+$ in the second half. 
Similarly, the third entry is a vector of length four, with each element corresponding to the number of information bits in one quarter of the codeword. 
This subdivision continues hierarchically until the $\log_2(N)$th entry, where each element corresponds to the number of information bits in a segment of length $N/2^{\log_2(N)} = 1$. 
The algorithm for computing $S\_values$ is presented in Algorithm~\ref{alg:calculate_s_values}.

The \textsc{Extract-Max} procedure maintains the heap property and extracts the element \(topS\), which corresponds to the path with the largest key (i.e., the highest path-metric value). 
The variable \(S_N\) denotes the number of paths in the stack. 
The bit index associated with the children of this best path is then stored as \(i\).

In line 18, using Algorithm~\ref{alg:updateLLR}, the \textsc{UpdateLLR} function outputs the updated intermediate LLRs \(\mathbf{L}\); the index vector for the intermediate LLRs corresponding to the special node, \(\mathbf{idx}_{\text{LLR}}\); the type of the special node, \(Type\); the index vector corresponding to the leaf nodes of the special intermediate node, \(\mathbf{idx}_{\text{chunk}}\); and the updated convolutional code's output vector, \(\mathbf{u}\).
Then, the intermediate LLRs corresponding to the special node of type \textit{Type} are stored in the vector \(\mathbf{r}_i\) of length \(N_{\text{chunk}}\).
In lines 21--31, using one of the four types of special nodes, the algorithm updates the stack.
The inputs to the special nodes are the stack \(S\);
the stack element corresponding to the best path, \(topS\); the number of paths in the stack, \(S_N\);
the state of the convolutional code, \(\mathbf{st}\); 
the intermediate LLR vector, \(\mathbf{L}\);
the convolutional code's output vector, \(\mathbf{u}\); 
the vector of bias values, \(\mathbf{E}_0\); 
and the indices: (i) the indices of the intermediate LLRs associated with the special node and (ii) the index vector corresponding to the leaf nodes of the special intermediate node. 
The outputs are the updated stack and the size of the stack.
Then the algorithm exits the while loop if decoding has reached level \(N\) of the decoding tree.


\begin{algorithm}[t!]
\small
\caption{$\textsc{f-Type-0}$ (Rate $0$ Node)}
\label{alg: f-type0}
\begin{algorithmic}[1]
\Input $\textit{topS}, S, S_N, \mathbf{st}, \mathbf{L}, \mathbf{idx}_{\text{LLR}}, 
        \mathbf{idx}_{\text{chunk}}, \mathbf{u}, \mathbf{c}, \mathbf{E}_0,$
\Statex \hspace{1.cm}$\text{S-Size}, \mathbf{r}_i, N_{\text{chunk}}$

\Output $S, S_N$

\For{$n \in \mathbf{idx}_{\text{chunk}}$}
    \State $\mathbf{u}[n] \gets \textsc{ConvEnc}(0, \mathbf{st}, \mathbf{c})$
    \State $\mathbf{st} \gets [0,~\mathbf{st}[1:\,|\mathbf{c}|-1]]$
\EndFor
\State $\mathbf{u}_{\text{temp}} \gets \textsc{PolarEncode}(\mathbf{u}[\mathbf{idx}_{\text{chunk}}])$
\State $\mathbf{m} \gets 1 - \log_2\!\big(1 + \exp(-\mathbf{r}_i \cdot (-1)^{\mathbf{u}_{\text{temp}}})\big) - \mathbf{E}_0[\mathbf{idx}_{\text{chunk}}]$
\If{$S_N = \text{S-Size}$}
    \State $[S, S_N] \gets \textsc{Extract-Min}(S, S_N)$
\EndIf
\State $S_N \gets S_N + 1$
\State $S[S_N].m \gets \textit{topS}.m + \text{sum}(\mathbf{m})$
\State $S[S_N].\mathbf{v} \gets [\textit{topS}.\mathbf{v},~ \mathbf{0}_{1\times N_{\text{chunk}}}]$
\State $S[S_N].\mathbf{st} \gets \mathbf{st}$ 
\State $S[S_N].\mathbf{L} \gets \mathbf{L}$ 
\State $S[S_N].\mathbf{u} \gets \mathbf{u}$
\State $S \gets \textsc{Heapify}(S, S_N)$
\State \Return $[S, S_N]$
\end{algorithmic}
\end{algorithm}

In Algorithm \ref{alg: f-type0}, the \textsc{f-Type-0} routine handles a type-0 (rate-0) special node, i.e., a block (chunk) in which all bits are frozen. 
Lines 1--4 run the one-bit convolutional encoder with input \(0\) over the indices in \(\mathbf{idx}_{\text{chunk}}\), writing the encoder output into \(\mathbf{u}[n]\) and advancing the convolutional state \(\mathbf{st}\) accordingly. 
At line 5, the algorithm forms the local partial sums by \(\mathbf{u}_{\text{temp}} \gets \textsc{PolarEncode}(\mathbf{u}[\mathbf{idx}_{\text{chunk}}])\) and computes the bit-metric vector
\(\mathbf{m}\),
which contains the contributions for the special-node bits.
If the stack is full (\(S_N=\text{S-Size}\)), \textsc{Extract-Min} evicts the worst path.
The algorithm then pushes a new path representing the zero decisions over this chunk: it updates the path metric by adding the metric of the partially explored path, appends \(N_{\text{chunk}}\) zeros to the path vector, stores the updated \(\mathbf{st}\), \(\mathbf{L}\), and \(\mathbf{u}\), and calls \textsc{Heapify} to preserve the heap property.


\begin{algorithm}[t!]
\small
\caption{$\textsc{f-Type-REP}$ (Rate $1/N_{\text{chunk}}$ Node)}
\label{alg: F-REP}
\begin{algorithmic}[1]
\Input $\textit{topS}, S, S_N, \mathbf{st}, \mathbf{L}, \mathbf{idx}_{\text{LLR}}, 
        \mathbf{idx}_{\text{chunk}}, \mathbf{u}, \mathbf{c}, \mathbf{E}_0,$
\Statex \hspace{1.cm}$\mathbf{RP}, \boldsymbol{\gamma}_T, \text{S-Size}, \mathbf{r}_i, N_{\text{chunk}}$

\Output $S, S_N$

\State $\mathbf{RP}_{\text{chunk}} \gets \mathbf{RP}[\mathbf{idx}_{\text{chunk}}]$

\State $p \gets \textsc{FindIndices}(\mathbf{RP}_{\text{chunk}}{=}1)$ \Comment{information position}

\State $\mathbf{V}_0, \mathbf{U}_0, \mathbf{U}_2 \gets \text{zeros}(2, N_{\text{chunk}})$
\State $\mathbf{V}_0[2][p] \gets 1$ \Comment{two candidates: all-zeros and single 1 at position $p$}
\State $\mathbf{ST} \gets \textsc{Replicate}(\mathbf{st}, 2)$

\For{$t = 1$ \textbf{to} $N_{\text{chunk}}$}
    \For{$j = 1$ \textbf{to} $2$}
        \State $\mathbf{U}_0[j][t] \gets \textsc{ConvEnc}(\mathbf{V}_0[j][t], \mathbf{ST}[j], \mathbf{c})$
    \EndFor
    \State $\mathbf{ST} \gets \big[ \mathbf{V}_0[:][t],~ \mathbf{ST}[:][1:|\mathbf{c}|-1] \big]$
\EndFor

\State $\gamma_{th} \gets \boldsymbol{\gamma}_T\big[\mathbf{idx}_{\text{chunk}}[p]\big]$

\For{$j = 1$ \textbf{to} $2$}
    \State $\mathbf{U}_2[j][:] \gets \textsc{PolarEncode}(\mathbf{U}_0[j][:])$
    \State $\mathbf{m} \gets 1 - \log_2\!\big(1 + \exp(-\mathbf{r}_i \cdot (-1)^{\mathbf{U}_2[j][:]})\big) - \mathbf{E}_0[\mathbf{idx}_{\text{chunk}}]$
    \If{$\mathbf{m}[p] > \gamma_{th}$}
        \If{$S_N = \text{S-Size}$}
            \State $[S, S_N] \gets \textsc{Extract-Min}(S, S_N)$
        \EndIf
        \State $S_N \gets S_N + 1$
        \State $S[S_N].m \gets \textit{topS}.m + \text{sum}(\mathbf{m})$
        \State $S[S_N].\mathbf{v} \gets [\textit{topS}.\mathbf{v}, ~~ \mathbf{V}_0[j][:]]$
        \State $S[S_N].\mathbf{st} \gets \mathbf{ST}[j]$
        \State $S[S_N].\mathbf{L} \gets \mathbf{L}$
        \State $\mathbf{u}[\mathbf{idx}_{\text{chunk}}] \gets \mathbf{U}_0[j][:]$
        \State $S[S_N].\mathbf{u} \gets \mathbf{u}$
        \State $S \gets \textsc{Heapify}(S, S_N)$
    \EndIf
\EndFor
\State \Return $[S, S_N]$
\end{algorithmic}
\end{algorithm}

In Algorithm \ref{alg: F-REP}, the \textsc{f-REP} routine implements the REP special node (rate \(1/N_{\text{chunk}}\)). 
It also supports an arbitrary placement of the information bit within the portion of the rate profile corresponding to this node.
Line 2 stores the position of the information bit in \(p\).
The matrices \(\mathbf{V}_0\), \(\mathbf{U}_0\), and \(\mathbf{U}_2\) correspond, respectively, to the two candidate inputs to the convolutional encoder, the corresponding convolutional outputs, and the intermediate partial sums at the layer of the REP node.
Line 5 replicates the convolutional state vector into \(\mathbf{ST}\) to track two candidates: the all-zeros vector and the vector with a single \(1\) at position \(p\).
Lines 6–11 compute, for both candidates, the convolutional outputs across the chunk and update the associated encoder states.
Line 12 extracts the bit-metric threshold \(\gamma_{th}\) at the information position from \(\boldsymbol{\gamma}_T\).

Lines 13–29 then, for each candidate \(j\in\{1,2\}\), form the partial sums via \(\textsc{PolarEncode}(\mathbf{U}_0[j][:])\), compute the bit-metric vector
\(\mathbf{m}\),
and compare \(\mathbf{m}[p]\) to \(\gamma_{th}\).
If \(\mathbf{m}[p] > \gamma_{th}\), the path is admitted to the stack (evicting the minimum element with \(\textsc{Extract-Min}\) if the stack is full), the path metric is increased, the local decisions \(\mathbf{V}_0[j][:]\) are appended, the updated state \(\mathbf{ST}[j]\) and buffers are stored, and \(\textsc{Heapify}\) restores the heap property.
Otherwise, the candidate is discarded.
Finally, the algorithm returns the updated stack \(S\) and its size \(S_N\).


\begin{algorithm}[t!]
\small
\caption{$\textsc{f-Type-IV}$ (Rate $2/N_{\text{chunk}}$ Node)}
\label{alg: process type 2}
\begin{algorithmic}[1]
\Input $\textit{topS}, S, S_N, \mathbf{st}, \mathbf{L}, \mathbf{idx}_{\text{LLR}}, 
        \mathbf{idx}_{\text{chunk}}, \mathbf{u}, \mathbf{c}, \mathbf{E}_0,$
\Statex \hspace{1.5cm}$\mathbf{RP}, \boldsymbol{\gamma}_T, \text{S-Size}, \mathbf{r}_i, N_{\text{chunk}}$

\Output $S, S_N$

\State $\mathbf{RP}_{\text{chunk}} \gets \mathbf{RP}[\mathbf{idx}_{\text{chunk}}]$

\State $\mathbf{p} \gets \textsc{FindIndices}(\mathbf{RP}_{\text{chunk}}{=}1)$ \Comment{two information positions}
\State $\mathbf{V}_0, \mathbf{U}_0, \mathbf{U}_2 \gets \text{zeros}(4, N_{\text{chunk}})$

\State $\mathbf{V}_0[1][:] \gets \mathbf{0}$ \Comment{Four candidate input patterns before convolution}
\State $\mathbf{V}_0[2][p_1] \gets 1$
\State $\mathbf{V}_0[3][p_2] \gets 1$
\State $\mathbf{V}_0[4][:] \gets \mathbf{RP}_{\text{chunk}}$

\State $\mathbf{ST} \gets \textsc{Replicate}(\mathbf{st}, 4)$

\For{$t = 1$ \textbf{to} $N_{\text{chunk}}$}
    \For{$j = 1$ \textbf{to} $4$}
        \State $\mathbf{U}_0[j][t] \gets \textsc{ConvEnc}(\mathbf{V}_0[j][t], \mathbf{ST}[j], \mathbf{c})$
    \EndFor
    \State $\mathbf{ST} \gets \big[ \mathbf{V}_0[:][t],~ \mathbf{ST}[:][1:|\mathbf{c}|-1] \big]$
\EndFor

\State $\gamma_{th} \gets \boldsymbol{\gamma}_T[\mathbf{idx}_{\text{chunk}}[p_1]] + \boldsymbol{\gamma}_T[\mathbf{idx}_{\text{chunk}}[p_2]]$

\For{$j = 1$ \textbf{to} $4$}
    \State $\mathbf{U}_2[j][:] \gets \textsc{PolarEncode}(\mathbf{U}_0[j][:])$
    \State $\mathbf{m} \gets 1 - \log_2\!\big(1 + \exp(-\mathbf{r}_i \cdot (-1)^{\mathbf{U}_2[j][:]})\big) - \mathbf{E}_0[\mathbf{idx}_{\text{chunk}}]$
    \If{$\mathbf{m}[p_1] + \mathbf{m}[p_2] > \gamma_{th}$}
        \If{$S_N = \text{S-Size}$}
            \State $[S, S_N] \gets \textsc{Extract-Min}(S, S_N)$
        \EndIf
        \State $S_N \gets S_N + 1$
        \State $S[S_N].m \gets \textit{topS}.m + \text{sum}(\mathbf{m})$
        \State $S[S_N].\mathbf{v} \gets [\textit{topS}.\mathbf{v}, ~~ \mathbf{V}_0[j][:]]$
        \State $S[S_N].\mathbf{st} \gets \mathbf{ST}[j]$
        \State $S[S_N].\mathbf{L} \gets \mathbf{L}$
        \State $\mathbf{u}[\mathbf{idx}_{\text{chunk}}] \gets \mathbf{U}_0[j][:]$
        \State $S[S_N].\mathbf{u} \gets \mathbf{u}$
        \State $S \gets \textsc{Heapify}(S, S_N)$
    \EndIf
\EndFor
\State \Return $[S, S_N]$
\end{algorithmic}
\end{algorithm}

In Algorithm \ref{alg: process type 2}, the \textsc{f-Type-IV} routine is an extension of \textsc{f-REP} that handles rate \(2/N_{\text{chunk}}\) nodes with two information bits. 
The information-bit positions \(p_1\) and \(p_2\) are obtained from the rate profile restricted to the current chunk. 
The algorithm evaluates four candidate input patterns: (i) all zeros, (ii) a single \(1\) at \(p_1\), (iii) a single \(1\) at \(p_2\), and (iv) ones at both \(\{p_1,p_2\}\) (i.e., \(\mathbf{RP}_{\text{chunk}}\)).

Compared with \textsc{f-REP}, the only substantive difference is the admission test: 
a candidate is kept if the \emph{sum} of the bit-metrics at the two information positions exceeds the \emph{sum} of their thresholds, i.e., \(\mathbf{m}[p_1]+\mathbf{m}[p_2]>\gamma_T[p_1]+\gamma_T[p_2]\); otherwise it is pruned. 
All other steps—convolutional encoding across the chunk, state tracking, forming partial sums via \textsc{PolarEncode}, heap maintenance through \textsc{Extract-Min} (when full) and \textsc{Heapify}, and the return values—proceed exactly as in \textsc{f-REP}. 
At the end, the algorithm returns the updated stack \(S\) and its size \(S_N\).


\begin{algorithm}[t!]
\scriptsize
\caption{$\textsc{f-Type-All}$ (Rate $N_{\text{chunk}}/N_{\text{chunk}}$ Node)}
\label{alg: f-type-all}
\begin{algorithmic}[1]
\Input $\textit{topS}, S, S_N, \mathbf{st}, \mathbf{L}, \mathbf{idx}_{\text{LLR}}, 
        \mathbf{idx}_{\text{chunk}}, \mathbf{u}, \mathbf{c}, \mathbf{E}_0,$
\Statex \hspace{1.5cm}$\boldsymbol{\gamma}_T, \text{S-Size}, \mathbf{r}_i, N_{\text{chunk}}$

\Output $S, S_N$

\State $A_{\text{Size}} \gets 2^{N_{\text{chunk}}}$

\State $A, C \gets \textsc{EmptyHeap}()$
\State $A[1].m \gets \textit{topS}.m$
\State $A[1].\mathbf{u}_d \gets \mathbf{0}_{1 \times N_{\text{chunk}}}$
\State $A[1].\text{Len} \gets 0$

\State $[A_N, \textit{count}, \text{row}_C]  \gets [1, 1, 0]$

\While{$\textit{count} < A_{\text{Size}}$}
    \State $[\textit{topA}, A, A_N] \gets \textsc{Extract-Max}(A, A_N)$
    \State $\mathbf{u}_d \gets \textit{topA}.\mathbf{u}_d$
    \State $L_{\text{current}} \gets \textit{topA}.\text{Len} + 1$

    \State $\gamma_{th} \gets \boldsymbol{\gamma}_T[\mathbf{idx}_{\text{chunk}}[L_{\text{current}}]]$

    \For{$b \in \{0,1\}$}
        \State \parbox[t]{.9\linewidth}{\scriptsize
$m \gets 1 - \log_2\!\big(1 + \exp\!\big(-\mathbf{r}_i[L_{\text{current}}] \cdot (-1)^{b}\big)\big) - \mathbf{E}_0[\mathbf{idx}_{\text{chunk}}[L_{\text{current}}]]$}

        \If{$m > \gamma_{th}$}
            \State $\mathbf{u}_d^{\text{temp}} \gets \mathbf{u}_d$
            \State $\mathbf{u}_d^{\text{temp}}[L_{\text{current}}] \gets b$

            \If{$L_{\text{current}} < N_{\text{chunk}}$}
                \If{$A_N = \text{S-Size}$}
                    \State $[A, A_N] \gets \textsc{Extract-Min}(A, A_N)$
                \EndIf
                \State $A_N \gets A_N + 1$
                \State $A[A_N].m \gets \textit{topA}.m + m$
                \State $A[A_N].\mathbf{u}_d \gets \mathbf{u}_d^{\text{temp}}$
                \State $A[A_N].\text{Len} \gets L_{\text{current}}$
                \State $A \gets \textsc{Heapify}(A, A_N)$
            \Else
                \If{$\text{row}_C = \text{S-Size}$}
                    \State $[C, \text{row}_C] \gets \textsc{Extract-Min}(C, \text{row}_C)$
                \EndIf
                \State $\text{row}_C \gets \text{row}_C + 1$
                \State $C[\text{row}_C].m \gets \textit{topA}.m + m$
                \State $C[\text{row}_C].\mathbf{u}_d \gets \mathbf{u}_d^{\text{temp}}$
                \State $C \gets \textsc{Heapify}(C, \text{row}_C)$
                \State $\textit{count} \gets \textit{count} + 1$
            \EndIf
        \Else
            \State $\textit{count} \gets \textit{count} + 2^{(N_{\text{chunk}} - L_{\text{current}})}$
        \EndIf
    \EndFor
\EndWhile

\For{$k = 1$ \textbf{to} $\text{row}_C$}
    \State $[\textit{topC}, C, \text{row}_C] \gets \textsc{Extract-Max}(C, \text{row}_C)$
    \State $\mathbf{U}_0 \gets \textsc{PolarEncode}(\textit{topC}.\mathbf{u}_d)$
    \State $\mathbf{V}_0 \gets \mathbf{0}_{1 \times N_{\text{chunk}}}$
    \State $\mathbf{st}^{\text{temp}} \gets \mathbf{st}$

    \For{$t = 1$ \textbf{to} $N_{\text{chunk}}$}
        \State $\mathbf{V}_0[t] \gets \textsc{ConvEncInv}(\mathbf{U}_0[t], \mathbf{st}^{\text{temp}}, \mathbf{c})$
        \State $\mathbf{st}^{\text{temp}} \gets [\,\mathbf{V}_0[t],~ \mathbf{st}^{\text{temp}}[1:|\mathbf{c}|-1]\,]$
    \EndFor

    \If{$S_N = \text{S-Size}$}
        \State $[S, S_N] \gets \textsc{Extract-Min}(S, S_N)$
    \EndIf

    \State $S_N \gets S_N + 1$
    \State $S[S_N].m \gets \textit{topC}.m$
    \State $S[S_N].\mathbf{v} \gets [\,\textit{topS}.\mathbf{v},~ \mathbf{V}_0\,]$
    \State $S[S_N].\mathbf{st} \gets \mathbf{st}^{\text{temp}}$
    \State $S[S_N].\mathbf{L} \gets \mathbf{L}$
    \State $\mathbf{u}^{\text{temp}} \gets \mathbf{u}$
    \State $\mathbf{u}^{\text{temp}}[\mathbf{idx}_{\text{chunk}}] \gets \mathbf{U}_0$
    \State $S[S_N].\mathbf{u} \gets \mathbf{u}^{\text{temp}}$
    \State $S \gets \textsc{Heapify}(S, S_N)$
\EndFor

\State \Return $[S, S_N]$

\end{algorithmic}
\end{algorithm}

In Algorithm \ref{alg: f-type-all}, the \textsc{f-Type-All} routine implements the rate-1 (all-information) special node. 
The maximum number of paths in the chunk subtree is \(A_{\text{Size}}\triangleq 2^{N_{\text{chunk}}}\).
We maintain two heaps: \(A\) for frontier (partial) paths and \(C\) for completed length-\(N_{\text{chunk}}\) paths.

Lines 7–40 implement a heap exploration of the subtree in which every bit is an information bit. 
At each expansion step, for the current position \(L_{\text{current}}\) and branch \(b\in\{0,1\}\), the bit-metric \(m\) is computed and compared against the corresponding threshold \(\gamma_{th}\).
If \(m>\gamma_{th}\), the child is admitted: if \(L_{\text{current}}<N_{\text{chunk}}\), it is pushed into \(A\) with updated metric and length; otherwise, the fully specified path is pushed into \(C\).
If \(m\le\gamma_{th}\), the branch is pruned and its remaining subtree is skipped by increasing \(\textit{count}\) by \(2^{(N_{\text{chunk}}-L_{\text{current}})}\).
In this algorithm, we assume a maximum stack size of $S_{\text{Size}}$ for both $A$ and $C$. When either heap is full, \textsc{Extract-Min} is used to evict the worst path before admitting a new one.

Lines 41–62 process completed candidates from \(C\) in descending metric order.
For each, \(\mathbf{U}_0\) is formed via \(\textsc{PolarEncode}(\cdot)\); the corresponding convolutional input \(\mathbf{V}_0\) and updated state \(\mathbf{st}^{\text{temp}}\) are recovered across the chunk using \textsc{ConvEncInv}, an inverse operation of the 1-bit convolutional code.
The resulting full-path is then appended to the global stack \(S\) (with \textsc{Extract-Min} if needed) and \textsc{Heapify} restores the heap property.
Finally, the algorithm returns the updated stack \(S\) and its size \(S_N\).


\begin{algorithm}[htpb]
\small
\caption{$\textsc{UpdateLLR}$}
\label{alg:updateLLR}
\begin{algorithmic}[1]
\Input $\mathbf{L}_{ch}, \mathbf{L}, i, \mathbf{u}, n, S\_values$
\Output $\mathbf{L}, \mathbf{idx}_{\text{LLR}}, \textit{Type}, \mathbf{idx}_{\text{chunk}}, \mathbf{u}$

\State $\mathbf{idx}_{\text{LLR}} \gets [\,]$
\State $N \gets 2^n$
\If{$i = 0$}
    \State $s \gets n - 1$ 
    \State $\text{sel} \gets 1$ \Comment{$f$ function}
\Else
    \State $s \gets \text{ffs}(i)$ \Comment{depth: $0,1,\ldots,n-1$ (0 is leaf)}
    \State $\text{sel} \gets 0$ \Comment{$g$ function}
\EndIf

\While{$s \ge 0$}
    \State $\text{two\_s} \gets 2^s$
    \If{$s = n-1$}
        \If{$i > 0 \land \text{sel} = 0$}
            \State $\mathbf{u}_n \gets \textsc{PolarEncode}\big(\mathbf{u}[\,i-\text{two\_s}+1 : i\,]\big)$
        \EndIf
        \State $\mathbf{L}^{(L)} \gets \mathbf{L}_{ch}[\,1:\text{two\_s}\,], $
        \State $\mathbf{L}^{(R)} \gets \mathbf{L}_{ch}[\,\text{two\_s}+1:2\text{two\_s}\,]$
        \If{$\text{sel} = 1$}
            \State $\mathbf{L}[\,1:\text{two\_s}\,] \gets f\!\big(\mathbf{L}^{(L)}, \mathbf{L}^{(R)}\big)$
        \Else
            \State $\mathbf{L}[\,1:\text{two\_s}\,] \gets g\!\big(\mathbf{L}^{(L)}, \mathbf{L}^{(R)}, \mathbf{u}_n\big)$
        \EndIf

        \State $\mathbf{idx}_{\text{LLR}} \gets [\,1:\text{two\_s}\,]$
        \State $\text{sel} \gets 1$
    \Else
        \State $\text{shift\_out} \gets N - 2^{s+1}$, 
        \State $\text{shift\_in} \gets N - 2^{s+2}$
        \If{$i > 0 \land \text{sel} = 0$}
            \State $\mathbf{u}_n \gets \textsc{PolarEncode}\big(\mathbf{u}[\,i-\text{two\_s}+1 : i\,]\big)$
        \EndIf
        \State $\text{shift\_L} \gets \text{shift\_in} + [\,1:\text{two\_s}\,]$
        \State $\text{shift\_R} \gets \text{shift\_in} + \text{two\_s} + [\,1:\text{two\_s}\,]$
        \State $\mathbf{idx}_{\text{LLR}} \gets \text{shift\_out} + [\,1:\text{two\_s}\,]$
        \If{$\text{sel} = 1$}
            \State $\mathbf{L}[\mathbf{idx}_{\text{LLR}}] \gets f\!\big(\mathbf{L}[\text{shift\_L}], \mathbf{L}[\text{shift\_R}]\big)$
        \Else
            \State $\mathbf{L}[\mathbf{idx}_{\text{LLR}}] \gets g\!\big(\mathbf{L}[\text{shift\_L}], \mathbf{L}[\text{shift\_R}], \mathbf{u}_n\big)$
        \EndIf
 
        \State $\text{sel} \gets 1$

        \State $\text{idx} \gets \left\lceil \dfrac{i+1}{\text{two\_s}} \right\rceil$
        \State $\text{level} \gets n - s + 1$
        \State $\textit{Type} \gets S\_values\{\text{level}\}(\text{idx})$
        \State $\text{chunk\_size} \gets \dfrac{N}{2^{\text{level}-1}}$

        \If{$\textit{Type} \in \{0, 1, 2, \text{chunk\_size}\}$}
            \State $\text{start} \gets (\text{idx}-1)\cdot \text{chunk\_size} + 1$
            \State $\text{end} \gets \text{idx}\cdot \text{chunk\_size}$
            \State $\mathbf{idx}_{\text{chunk}} \gets [\,\text{start}:\text{end}\,]$
            \State \textbf{break}
        \EndIf
    \EndIf
    \State $s \gets s - 1$
\EndWhile

\State \Return $\big[\mathbf{L}, \mathbf{idx}_{\text{LLR}}, \textit{Type}, \mathbf{idx}_{\text{chunk}}, \mathbf{u}\big]$
\end{algorithmic}
\end{algorithm}

Algorithm~\ref{alg:updateLLR} updates the intermediate LLRs.
The inputs are the channel LLRs \(\mathbf{L}_{ch}\), the intermediate LLRs \(\mathbf{L}\), the running path bit index \(i\), the convolutional code output vector \(\mathbf{u}\), the tree depth \(n=\log_2 N\), and the precomputed \(S\_values\) (Alg.~\ref{alg:calculate_s_values}).
The outputs are the updated \(\mathbf{L}\); \(\mathbf{idx}_{\text{LLR}}\), the index vector for the intermediate LLRs corresponding to the special node; the type of the special node; \(\mathbf{idx}_{\text{chunk}}\), the index vector corresponding to the leaf nodes of the special intermediate node; and the updated vector \(\mathbf{u}\).
The LLRs can be updated as~[1]
\begin{equation}
\begin{aligned}
& f(\mathbf{L}^{(L)},\mathbf{L}^{(R)}) = 2\,\tanh^{-1}\!\Big(\tanh\!\big(\tfrac{\mathbf{L}^{(L)}}{2}\big)\,\tanh\!\big(\tfrac{\mathbf{L}^{(R)}}{2}\big)\Big),\\
& g(\mathbf{L}^{(L)},\mathbf{L}^{(R)}, \mathbf{u}_n) = \mathbf{L}^{(L)}\,(-1)^{\mathbf{u}_n} + \mathbf{L}^{(R)}.
\end{aligned}
\end{equation}

The variable \(s\in\{0,1,\ldots,n-1\}\) denotes the depth in the polarization tree (with \(s=0\) at the leaves).
The variable \(\text{sel}\) determines whether the \(f\) or \(g\) function is used: \(\text{sel}=0\) uses \(g\), and \(\text{sel}=1\) uses \(f\).
If \(i=0\), the algorithm starts from the root (\(s\gets n-1\)) and uses the \(f\) function first (\(\text{sel}\gets 1\)).
Otherwise, the depth is determined by find-first-bit-set (ffs) operation introduced in \cite{leroux2012hardware}.
To apply the \(g\) function, we compute the checksum \(\mathbf{u}_n\) at depth \(s\) by polar-encoding the segment \(\mathbf{u}[\,i-2^s+1 : i\,]\) (lines 14 and 29).
The algorithm then applies \(f\) or \(g\) to the parent-node LLRs: in lines 12–24 the parent LLRs come from the channel, and in lines 26–38 from the intermediate LLRs \(\mathbf{L}\).
The selector is flipped to \(f\) (\(\text{sel}\gets 1\)) for the next lower stage.

After updating the LLRs at a depth of the polarization tree, the algorithm locates the \emph{chunk} that contains index \(i\) and obtains, from the precomputed partition summary \(S\_values\), the number of information bits in this chunk (line 42).
If \(\textit{Type}\in\{0,1,2,\text{chunk\_size}\}\), the algorithm identifies a special node and exits the loop early.
This enables the caller (the fast stack decoder) to switch to a node-specialized routine for the detected chunk.
At each visited stage, the algorithm records the updated LLR block in \(\mathbf{idx}_{\text{LLR}}\), ensuring the caller knows exactly which entries of \(\mathbf{L}\) changed.
It finally returns \(\big(\mathbf{L},\mathbf{idx}_{\text{LLR}},\textit{Type},\mathbf{idx}_{\text{chunk}},\mathbf{u}\big)\).


\begin{algorithm}[htpb]
\small
\caption{$\textsc{Calculate-S-Values}$}
\label{alg:calculate_s_values}
\begin{algorithmic}[1]
\Input $\mathbf{RP}$
\Output $S\_values$

\State $N \gets \text{length}(\mathbf{RP})$
\State $\text{max\_depth} \gets \log_2(N)$
\State $S\_values \gets \text{cell}(\text{max\_depth}+1, 1)$

\State $S\_values\{1\} \gets \text{sum}( \mathbf{RP})$ \Comment{root node at depth 0}

\For{$d = 1 \to \text{max\_depth}$}
    \State $\text{num\_nodes} \gets 2^d$
    \State $S\_values\{d+1\} \gets \mathbf{0}_{1 \times \text{num\_nodes}}$
    \For{$i = 0 \to \text{num\_nodes}-1$}
        \State $\text{start} \gets i \cdot \dfrac{N}{\text{num\_nodes}} + 1$
        \State $\text{end} \gets (i+1) \cdot \dfrac{N}{\text{num\_nodes}}$
        \State $S\_values\{d+1\}[i+1] \gets \text{sum}( \mathbf{RP}[\text{start}:\text{end}])$
    \EndFor
\EndFor

\State \Return $S\_values$
\end{algorithmic}
\end{algorithm}


In Algorithm~\ref{alg:calculate_s_values}, the procedure computes the hierarchical distribution of information bits across different segments of the codeword based on the given rate profile $\mathbf{RP}$. 
The algorithm constructs $S\_values$ as a cell array with $\log_2(N)+1$ entries, where $N$ is the codeword length.
The first entry, $S\_values\{1\}$, stores the total number of information bits $K$, obtained by summing all elements of $\mathbf{RP}$. 
For each subsequent depth level $d = 1, 2, \dots, \log_2(N)$, the algorithm divides the codeword into $2^d$ equally sized segments and computes the number of information bits within each segment. 
Thus, $S\_values\{d+1\}$ is a vector of length $2^d$, where the $i$-th element corresponds to the number of information bits in the $i$-th segment of length $N/2^d$. 
This hierarchical representation captures the distribution of information bits from the entire codeword down to individual bit positions.




\section{Numerical Results}\label{sec:Numerical}
In this section, we evaluate the performance of our proposed decoding schemes over a BI-AWGN channel with BPSK modulation. 
To define the threshold parameter, one may use a theoretical bound, such as the RCU bound \cite{polyanskiy2010channel}, denoted by \(D\), and set the threshold parameter to \(P_{\mathrm{th}} = D/10\).
Since, for the short block lengths considered, a Fano decoder with FER \(D\) at a given \(E_b/N_0\) achieves performance very close to the theoretical bounds, we also can set the threshold parameter to \(P_{\mathrm{th}} = D/10\) in our numerical results.
The division by 10 is used to impose a more conservative performance limit.
For all evaluations, we adopt the connection polynomial $\mathbf{c}(t) = t^{10} + t^9 + t^7 + t^3 + 1$ \cite{moradi2020performance}. 
Furthermore, for Fano decoding, we fix the threshold spacing to $\Delta = 2$.

\begin{figure}[htbp]
\centering
	\includegraphics [width = \columnwidth]{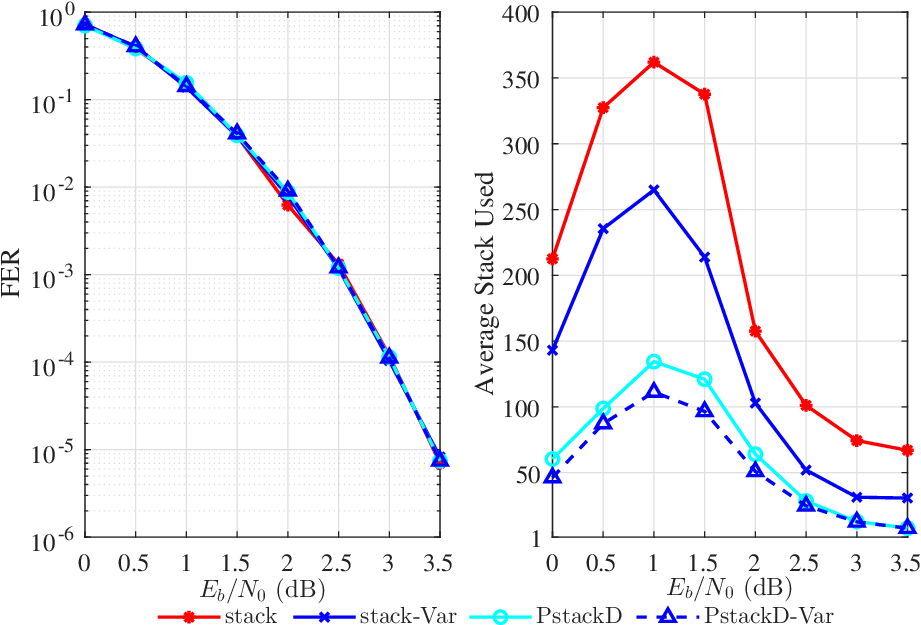}
	\caption{FER performance and the average number of paths in the stack comparison of PAC$(128,64)$ codes with different path-pruning strategies for the stack decoding algorithm. } 
	\label{fig:ANSg_PstackD_Var}
\end{figure}

Fig. \ref{fig:ANSg_PstackD_Var} compares the performance of the conventional stack decoding algorithm for a PAC$(128,64)$ code with an RM-based rate profile with that of the path-pruning strategy in stack decoding (PstackD) proposed in \cite{moradi2023tree}, which is based on the Chernoff bound, and our proposed path-pruning strategy based on bit-channel variances (PstackD-Var), which uses both the Chernoff and Chebyshev bounds. It also compares these schemes with the pruning strategy based only on the variance (Stack-Var), which relies only on the Chebyshev bound. The average number of paths in the stack (average stack used) is also shown in this figure.

At an $E_b/N_0$ value of $1~$dB, our proposed pruning method achieves a $70\%$ reduction in the average number of paths in the stack compared to the conventional stack decoding algorithm, and an $18\%$ reduction compared to the method in \cite{moradi2023tree}.


\begin{table*}[t]
\centering
\caption{Average Performance Metrics of PstackD-Var Decoder for PAC$(64,57)$ (Stack Size = $8$, Max Cycles = $32$)}
\label{tab:PstackDVar_PAC64_57}
\begin{tabular}{|c||c|c|c|c|c|c|c|c|c|c|c|}
\hline
$E_b/N_0$ (dB) & 2.0 & 2.5 & 3.0 & 3.5 & 4.0 & 4.5 & 5.0 & 5.5 & 6.0 & 6.5 & 7.0 \\ \hline\hline
Average Cycles     & 16.04 & 16.59 & 16.71 & 16.29 & 15.72 & 15.35 & 15.14 & 15.04 & 15.01 & 15.00 & 15.00 \\ \hline
Average Stack Used &  6.68 &  6.50 &  6.03 &  5.41 &  4.76 &  4.12 &  3.84 &  3.54 &  3.21 &  2.90 &  2.76 \\ \hline
Average $f$ \& $g$ Operations & 30.25 & 31.35 & 31.55 & 30.66 & 29.48 & 28.72 & 28.28 & 28.09 & 28.02 & 28.00 & 28.00 \\ \hline
\end{tabular}
\end{table*}


\begin{table*}[t]
\centering
\caption{Average Performance Metrics of PstackD-Var Decoder for PAC$(128,99)$ (Stack Size = $64$, Max Cycles = $1024$)}
\label{tab:PstackDVar_PAC128_99}
\begin{tabular}{|c||c|c|c|c|c|c|c|c|c|c|c|}
\hline
$E_b/N_0$ (dB) & 0.0 & 0.5 & 1.0 & 1.5 & 2.0 & 2.5 & 3.0 & 3.5 & 4.0 & 4.5 & 5.0 \\ \hline\hline
Average Cycles     & 37.87 & 42.02 & 53.47 & 81.01 & 110.96 & 95.38 & 65.29 & 44.88 & 37.47 & 35.47 & 35.08 \\ \hline
Average Stack Used & 24.63 & 22.86 & 23.98 & 29.02 & 31.79 & 26.05 & 19.44 & 14.29 & 11.63 & 10.65 &  9.69 \\ \hline
Average $f$ \& $g$ Operations & 74.11 & 82.66 & 106.05 & 161.76 & 222.32 & 191.15 & 130.12 & 88.30 & 73.08 & 68.96 & 68.16 \\ \hline
\end{tabular}
\end{table*}


\begin{figure}[htbp]
\centering
	\includegraphics[width = \columnwidth]{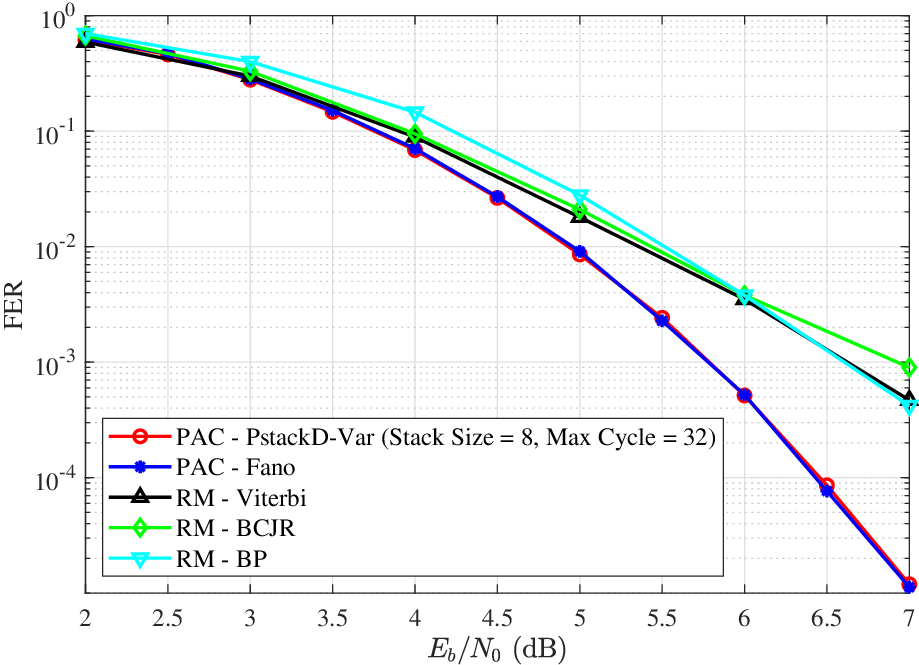}
	\caption{FER performance comparison of PAC$(64,57)$ and RM$(64,57)$ codes under different decoding algorithms.}
	\label{fig:VarStack_Fano_N64_K57}
\end{figure}

Fig.~\ref{fig:VarStack_Fano_N64_K57} illustrates the FER performance of the PAC$(64,57)$ code with an RM-based rate profile decoded using our proposed fast stack decoding algorithm with a stack size of $8$ and a maximum of $32$ decoding cycles, compared against Fano decoding, where no limit is imposed on the number of explored paths. 
For comparison, we also evaluate the FER performance of the corresponding RM$(64,57)$ code under Viterbi, Bahl–Cocke–Jelinek–Raviv (BCJR), and belief-propagation (BP) decoding as in \cite{arikan2009performance}. 
It can be observed that while Viterbi and BCJR decoding provide near-maximum-likelihood (ML) performance, their computational complexity grows exponentially with the number of frozen bits, making them impractical for larger block lengths.

Table~\ref{tab:PstackDVar_PAC64_57} also reports the average number of decoding cycles, the average stack size used, and the average number of times the $f$ and $g$ functions are invoked. 
For reference, in conventional successive-cancellation (SC) decoding of polar codes, the $f$ and $g$ functions are called exactly $2N-2 = 126$ times for a code length of $N=64$. 
In contrast, for stack decoding, multiple candidate paths are maintained simultaneously, and the reported averages in the table are computed per decoded codeword, representing the total number of $f$ and $g$ function calls accumulated across all explored paths during the decoding process.


\begin{figure}[htbp]
\centering
	\includegraphics[width = \columnwidth]{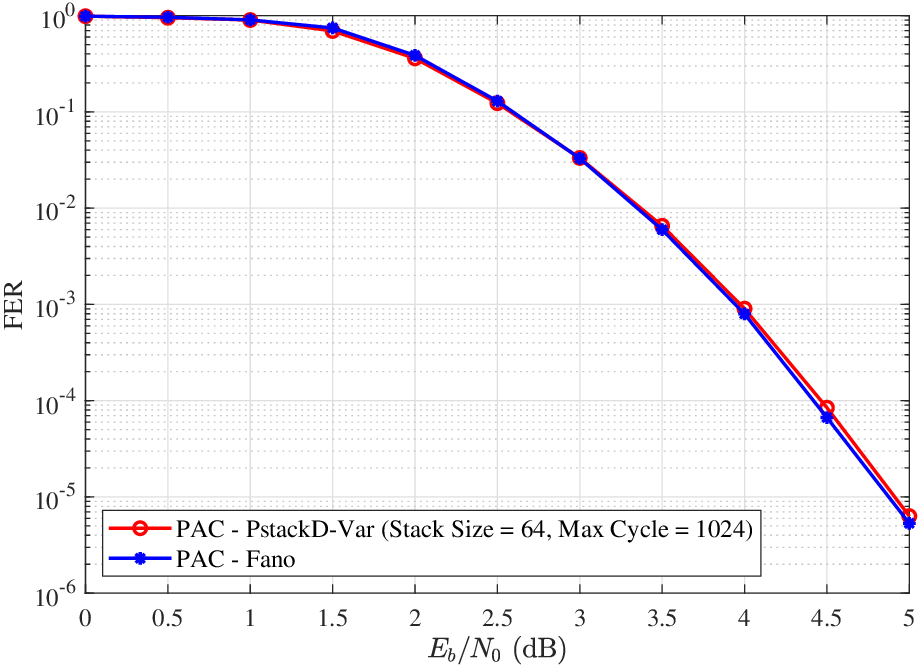}
	\caption{FER performance comparison of PAC$(128,99)$ codes with Fano and stack decoding algorithms.}
	\label{fig:PAC_Stack_vs_Fano_K99N128}
\end{figure}

Fig.~\ref{fig:PAC_Stack_vs_Fano_K99N128} illustrates the FER performance of the
PAC$(128,99)$ code with an RM-based rate profile decoded uing our
proposed stack decoding algorithm with a stack size of $64$ and a maximum of $1024$ decoding cycles. 
The results demonstrate that the proposed approach achieves FER performance comparable to that of the Fano sequential decoding algorithm.

Table~\ref{tab:PstackDVar_PAC128_99} also presents the average number of decoding cycles, the average stack size used, and the average number of times the $f$ and $g$ functions are invoked for the PAC$(128,99)$ code. 
For reference, in conventional SC decoding of polar codes, the $f$ and $g$ functions are called exactly $2N-2 = 254$ times for a code length of $N=128$.




\begin{figure}[htbp]
\centering
	\includegraphics[width = \columnwidth]{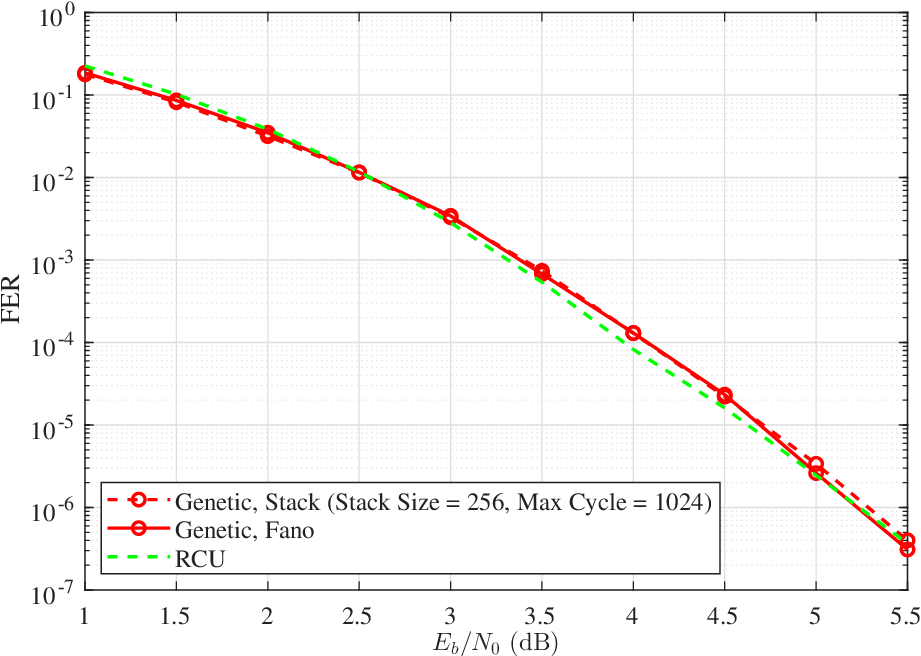}
	\caption{FER performance comparison of PAC$(64,32)$ codes with Fano and stack decoding algorithms.}
	\label{fig:PAC_Stack_vs_Fano_K32N64}
\end{figure}

Fig.~\ref{fig:PAC_Stack_vs_Fano_K32N64} illustrates the FER performance of a PAC$(64,32)$ code whose rate profile is obtained by a genetic-algorithm-based optimization initialized from the RM rate profile \cite{moradi2024pac_Genetic}. The code is decoded using our proposed stack decoding algorithm with a stack size of $256$ and a maximum of $1024$ decoding cycles, and its performance is compared with that of Fano decoding without any limitation on the search effort.
The results demonstrate that the proposed approach achieves FER performance comparable to that of the Fano sequential decoding algorithm.


\begin{table*}[t]
\centering
\caption{Average Performance Metrics of PstackD-Var Decoder for PAC$(64,32)$ (Stack Size = $256$, Max Cycles = $1024$)}
\label{tab:PstackDVar_PAC64_32}
\begin{tabular}{|c||c|c|c|c|c|c|c|c|c|c|}
\hline
$E_b/N_0$ (dB) & 1.0 & 1.5 & 2.0 & 2.5 & 3.0 & 3.5 & 4.0 & 4.5 & 5.0 & 5.5 \\ \hline\hline
Average Cycles     & 63.57 & 51.13 & 33.73 & 25.04 & 20.02 & 17.52 & 16.61 & 16.27 & 16.13 & 16.07 \\ \hline
Average Stack Used & 78.81 & 60.27 & 44.54 & 33.79 & 26.51 & 23.01 & 20.93 & 20.09 & 19.25 & 18.82 \\ \hline
Average $f$ \& $g$ Operations & 113.08 & 91.77 & 61.21 & 45.96 & 37.06 & 32.59 & 30.97 & 30.39 & 30.18 & 30.09 \\ \hline
\end{tabular}
\end{table*}

Table~\ref{tab:PstackDVar_PAC64_32} also presents the average number of decoding cycles, the average stack size used, and the average number of times the $f$ and $g$ functions are invoked for the PAC$(64,32)$ code.
As expected, these complexity measures decrease as $E_b/N_0$ increases.
For reference, in conventional SC decoding of polar codes, the $f$ and $g$ functions are called exactly $2N-2 = 126$ times for a code length of $N=64$.

\section{Conclusion}\label{sec: Conclusion}

In this paper, we introduced a fast stack decoding algorithm for PAC codes that efficiently handles rate-0, REP, type-IV, and rate-1 special nodes. 
For a rate-1 node with $N_0$ leaf nodes in the polarization tree, a conventional stack decoder may need to explore up to $2^{N_0}$ candidate paths, which becomes prohibitively complex for high-rate codes. 
In contrast, unlike the fast list decoding algorithm that restricts exploration to only $2L$ paths for a list size of $L$, the proposed approach addresses the computational challenge of rate-1 nodes for the fast stack decoding more effectively. 
To tackle this issue, we leverage the polarization properties of the bit-metric function and introduce a novel pruning criterion that selectively discards low-probability paths while ensuring that the probability of pruning the correct path decreases exponentially. 
Furthermore, we propose an accurate approximation method for estimating the polarized varentropy over BI-AWGN channels, enabling the design of adaptive and efficient pruning thresholds.


\ifCLASSOPTIONcaptionsoff
  \newpage
\fi

\bibliographystyle{IEEEtran}
\bibliography{bibliography}

\end{document}